\documentclass[twocolumn,showpacs,amsmath,amssymb]{revtex4}
\usepackage{graphicx}
\usepackage{dcolumn}
\usepackage{bm}

\newcommand{\ba}{\begin{eqnarray}}
\newcommand{\ea}{\end{eqnarray}}
\newcommand{\be}{\begin{equation}}
\newcommand{\ee}{\end{equation}}
\newcommand{\bd}{\begin{displaymath}}
\newcommand{\ed}{\end{displaymath}}

\begin{document}

\title{Spectral Broadening of Radiation  from Relativistic
Collapsing Objects}
\author{Valeri P. Frolov$^*$, Kyungmin Kim$^\dagger$ and  Hyun Kyu
Lee$^*,^\dagger$}
\address{
  \medskip
  $^*$Theoretical Physics Institute, Department of Physics, University of
Alberta\\
  Edmonton, AB, Canada, T6G 2J1\\
  {\rm E-mail: \texttt{frolov@phys.ualberta.ca}}
  \medskip
}
\address{
  \medskip
  $^\dagger$Department of Physics and BK21 Division of Advanced Research and
Education in Physics,
   Hanyang University, Seoul
  133-791, Korea\\
  {\rm E-mail: \texttt{hyunkyu@hanyang.ac.kr}}
  \medskip
}
\date{\today}


\begin{abstract} We  discuss light curves and the spectral broadening
of the radiation emitted during the finite interval of time from the
surface of a spherically symmetric collapsing object. We study a
simplified model of monochromatic radiations.  We discuss how one
can obtain information about the physical parameters of the
collapsing body, such as its mass and radius, from the light curves
and spectral broadenings.
\end{abstract}

\pacs{PACS numbers:  04.40.Dg, 95.85.-e,97.60.-s \hfill
Alberta-Thy-02-07}
\maketitle

Light propagating in the vicinity of astrophysical compact
objects, like neutron stars and black holes, is affected by the
gravitational field. It has been demonstrated that the general
relativistic effects might be important for understanding the
features of the radiation coming from the neutron star like
objects \cite{pfc,ll,bel}. The gravitational redshift and bending
of light rays  emitted by a compact object affect  the form and
spectrum of the observed signals. For the emission of light from
the vicinity of a black hole these effects are more profound than
for neutron stars. For example, the line broadening of X-ray
observed by the ASCA satellite can be explained by the strong
gravitational effect on the light emitted from the accretion disk
located near to the central black hole \cite{asca,BA}.

Astrophysical black holes are believed to be formed as a result of
the gravitational collapse of massive stars \cite{nsc}.  Then it
is natural to expect that the  radiation emitted during the
gravitational collapse is also affected by the strong gravity. For
a spherical collapse and continuous emission of light this effect
was studied in details \cite{at,jaffe,lr}.  Recently the light
curves for collapsing objects were studied in a slightly different
set up \cite{fl} assuming that the radiation has a profile of a
sharp  in time pulse. Such radiation may occur during the collapse
of a star when its matter density becomes much higher than the
nuclear density. Under these conditions hadronic phase transitions
are expected \cite{schafer} which may result in sharp-in-time
emission of massless particles (photons and neutrino) \cite{lp2}.

In this work,  we consider the radiation emitted by a collapsing star
during a finite time interval and
calculate light curves and the spectrum of this radiation as seen by a
distant observer. As in the previous work \cite{fl}, we adopt a
simplified model of  a freely falling spherical surface and assume
that the radiation is originally monochromatic.  But instead of
instant radiation, we focus on  the radiation emitted during the
finite interval of time. The main goal of this study is to analyze
how one can extract information about the characteristics of a
collapsing object (its mass and radius) from the observed
spectra and light curves.

In the adopted simplified approach, we consider a spherical
collapse  and assume that the points of the collapsing surface
follows  radial geodesics in the Schwarzschild geometry
\cite{at,fl,os,sha}. Thus the photons emitted from the surface
propagate to the observer at infinity in the background of the
Schwarzschild metric
    \ba
    ds^2 = -f dt^2 + f^{-1} dr^2 + r^2 d\Omega^2
    \label{schw}\, .
    \ea
Here $f=f(r) = 1- 2M/r$,   $M$ is the mass of the collapsing
object, and $d\Omega^2=d\theta^2+\sin^2\theta\, d\phi^2$ is a line
element on a unit sphere. We use the natural units, $c=G= \hbar
=1$ throughout this paper.

Denote by $\tau$ the proper time as measured by an observer comoving
with the collapsing surface. We suppose that the collapse starts  at
$\tau=0$ with the initial surface radius $R_0$.  We denote  by
$t^{(e)}(\tau)$ the Schwarzschild time $t$ corresponding to $\tau$
and take $t^{(e)}(\tau=0)=0$. In the coordinates $(t,r,\theta,\phi)$
the  four-velocity of the collapsing  surface is $ v^{\mu} =
(dt^{(e)}/d\tau, dR/d\tau, 0, 0)$. We define  $v_I$ as the invariant
radial velocity which measures the proper length change as measured
by the proper time of the observer at rest at a given radius. One has
$v_I=f^{-1}(R)\frac{dR}{dt}\,$, $|v_I|\leq 1$.  For a freely falling
surface with the initial radius $R_0$ the invariant
velocity  as a function of  $R$  is

\ba v_I=-\sqrt{\frac{2M}{R}}
     \frac{\sqrt{1-R/R_0}}{\sqrt{1-2M/R_0}}\, . \label{v_s}
\ea

Because of the spherical symmetry the trajectories of particles
and light are plane, so that we can always put $\phi=0$. Denote by
$p^{\mu} = (p^t, p^r, p^{\theta}, 0)$ the 4-momentum of a photon,
then   $E=-p_t$ (the energy at infinity) and $L=p_{\theta}$ (the
angular momentum) are constants of motion. In the adopted units,
where $\hbar=1$, $E$ coincides with the photon frequency as
measured by a far distant observer, and ${\bf p}$ coincides with
its wavevector. Instead of the angular momentum $L$ we shall use
the impact parameter defined by $ l = L/E\,$. The radial momentum
$p^r$ is given by $ p^r = \sigma E Z$,  where $ Z(l,r)=\sqrt{1 -
l^2 f(r)/r^2}\,$ .  Here and later $\sigma$ denotes a sign
function which takes the values $+$ and $-$ for a forward
($p_r>0$) and backward ($p_r<0$) motion of the photon,
respectively.

For a photon emitted at the radius $R$ and propagating to the
infinity there exists  an upper limit for the impact parameter
$l_{max} = R/\sqrt{f(R)}$ determined by the condition
$Z(l_{max},R)=0$. Let us consider the emission angle, $\beta$, of
the radiation as measured by an observer comoving with the
surface. The emission angle is zero for the emission in vertical
outward direction and $\pi/2$ for the tangentially emitted
radiation. Since the matter inside the collapsing surface is
supposed to be opaque, it is natural to consider only the light
emitted with the emission angle, $\beta \leq \pi/2$. It gives a
lower bound on the impact parameter for the backward emission,
$l_T  \leq l \leq l_{max}$, where $l_T =R/\sqrt{f(R_0)}$ is
determined by the condition for tangentially emitted radiation,  $
Z(l_{T},R)=-v_I\,$. Hence the possible ranges of the impact
parameter are $0 \leq l \leq l_{max}$ and $l_T \leq l \leq
l_{max}$ for a forward and backward emission, respectively. In
this work we consider only $R>3\sqrt{3}\sqrt{1- 2M/R_0}M$. A
discussion of the allowed ranges of the impact parameter for the
smaller radius up to $R \sim 2M$, can be found in \cite{at,jaffe}.

We choose the direction of the axis $\theta=0$ such that a photon
emitted with $\beta=0$ at $\theta=0$ propagates along the radial
direction to a distant observer. For a photon  emitted from a
collapsing surface  at the angle $\theta$ to reach the distant
observer, $\theta$ coincides with the bending angle. For a null
ray from the collapsing surface when its radius is $R$, the
bending angle for a forward-emission, $\theta_+$  is
      \ba \theta_+ = \Theta(l,R)\equiv l\int^{\infty}_{R}
\frac{dr}{r^2 Z(l,r) }
\label{Theta}\, .
    \ea
For a  backward-emission,  a photon, before it reaches the infinity,
should first pass through a turning point $r_t (< R)$.  The turning
point is determined by $Z(l,r_t)=0$. Then we get  the bending angle
for backward emission as $ \theta_{-}(l,R)
=2\Theta(l,r_t)-\Theta(l,R)$. One can easily check that for a
tangentially emitted backward photon the angle $\theta_-(l_T, R)$ is
greater than $\pi/2$. This means that a distant observer can see
a part of the `opposite side' of the spherical surface as a result
of strong gravity effect.

Let $p_{\mu}^{(e)}$ and $p_{\mu}^{(o)}$ be 4-momentum of a photon
emitted by a collapsing surface and of a photon  at infinity
respectively. Then $\nu^{(e)}=-p_{\mu}^{(e)}v^{\mu}$ is the energy
(frequency) of the photon as measured by a comoving observer. For
the observer at rest at with the 4-velocity,
$v^{(o)\mu}=\delta^{\mu}_0$, the observed energy (frequency) is
determined by $\nu^{(o)}=-p_{\mu}^{(o)}v^{(o)\mu}$.   For a given
ray with the impact parameter $l$, which is emitted from the
freely-falling surface, Eq. (\ref{v_s}), when its radius is $R$, the
redshift factor $\Phi$ defined as the ratio of the emitted frequency
to the observed frequency  at infinity, $\Phi \equiv
\nu^{(e)}/\nu^{(o)}\,$, is given by \ba \Phi_{\sigma}(l,R) = {1
-\sigma v_I Z(l,R)\over \sqrt{f}\sqrt{1-v_I^2}}\, .\label{z} \ea

Consider a light ray with the impact parameter $l$ emitted from the
collapsing surface at the moment of the proper time  $\tau$, when it
has the radius $R(\tau), R_f(\tau_f) \leq R(\tau) \leq R_i(\tau_i)$,
and let $t$ be the time when it reaches the distant observer. The
first ray reaching the distant observer is the ray with $l=0$
emitted from  a point `a'  at $\tau_i$ (see Fig. 1). We use it as a
reference ray. We characterize the arrival time of other rays by
their time delay $\Delta t$ with respect to the arrival time of the
reference ray. In Fig. 1  we show the sets of points of emission $\{
R,\theta \}$, which have the same $\Delta t$,  by dotted lines. For
the forward ray the time delay $\Delta t$  is given by the following
expression \ba
    \Delta t_{+}(l;\tau,\tau_i)
     = t^{(e)}(\tau)-t^{(e)}(\tau_i)  +T(l,R(\tau))
\nonumber\\
+ R_i -R(\tau)
    + 2M \ln
    \frac{R_i-2M}{R(\tau) - 2M}\label{dtflr},\\
T(l,R) \equiv \int^{\infty}_{R} \frac{dr}{f(r)}
\left[\frac{1}{Z(l,r)}-1\right]\,. \label{TT}
     \ea
Similarly for the  backward ray one has
\ba  \label{dtblr}
    \Delta t_{-}(l;\tau,\tau_i)
    =  t^{(e)}(\tau)-t^{(e)}(\tau_i)
    + 2T(l,r_t) -T(l,R(\tau))
\nonumber
{ }\hspace{-0.5cm}\\
+  R_i + R(\tau) - 2 r_t + 2M \ln
    \frac{(R(\tau)-2M)(R_i-2M)}{(r_t -2M)^2} .
\ea

The integrals for $\Theta$ and $T$, Eqs. (\ref{Theta}) and
(\ref{TT}) respectively, can be expressed in terms of the elliptic
functions. However, for practical calculations it is very convenient
to use the analytical  approximations for these quantities in terms
of simple elementary functions \cite{fl,cf}.

\begin{figure}[htp]
\begin{center}
\includegraphics[height=1.92in]{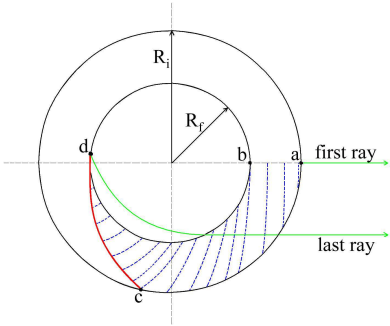}
\caption{ The dotted lines are the emission points for the same
arrival time parameter.`a' and `b' denote the points of  the first
lights with $l=0$ from $R_i$ and $R_f$ respectively.  The solid
line  from  `c' to `d' are the points where the last tangential
ray for each $R$, $R_i \geq R \geq R_f $, is emitted.  `c' denotes
where the light is emitted tangentially from $R_i$. `d' is the
point from which the last ray is emitted from $R_f$. }
\label{bangle}
\end{center}
\end{figure}

For the monochromatic emission with frequency $\nu^{(e)}=\nu$, the
flux as measured by a distant observer at $r_0$  can be written as
an integral over the proper time, $\tau$: \ba F_{\nu}^{(o)}(t)=
\frac{2\pi}{r_0^2} \int d\tau  l \left|dl\over d\tau \right|
\Phi^{-4}  I_{\nu}^{(e)}(l,\tau) \, , \label{fnu} \ea where $
I_{\nu}^{(e)}(l,\tau)$ is the intensity of the radiation (see e.g.
\cite{fl}). In the present paper we focus on the finite-duration
monochromatic emission, and take  the intensity in the form $
I_{\nu}^{(e)}(l,\tau) = f(\tau){\cal I}^{(e)}(l)\, .$ For simplicity
we assume that the profile $f(\tau)$ is  a step function, $ f(\tau)
= 1$  for $\tau_i \leq \tau \leq\tau_f$ and $ f(\tau) = 0$
otherwise. We further simplify calculations assuming the isotropic
emission: ${\cal I}^{(e)}(l)={\cal I}^{(e)}$. Then we get \ba
F_{\nu}^{(o)}(t)={2\pi{\cal I}^{(e)}\over r_0^2} \int d \tau f(\tau)
l \left|dl\over d\tau \right| \Phi^{-4}\, . \label{fot} \ea  The
integration over $\tau$ for a given $t$ is equivalent to the
integration along the  dotted line in Fig. \ref{bangle} for the same
arrival time difference. The duration of the observed light, $\Delta
T$, is determined by the difference between the arrival time of the
last ray from $R_f$ emitted at the point `d'  and the arrival time
of the first ray from $R_i$ emitted at `a' with $l=0$: \ba
\label{duration}
    \Delta T &\equiv&\Delta t_{-}(l_T,\tau_f,\tau_i)
    =  t^{(e)}(\tau_f)-t^{(e)}(\tau_i)\nonumber\\
    &+& 2T(l_T,r_t) -T(l_T,R(\tau))+  R_i + R(\tau_f)\\
&-& 2 r_t + 2M \ln
    \frac{(R(\tau_f)-2M)(R_i-2M)}{(r_t -2M)^2} .\nonumber
\ea
  In what follows it is convenient to  use a dimensionless time
parameter $\delta$, which is  a normalized arrival time difference
defined as $\delta  \equiv \Delta t_{\pm} / \Delta T$. The time
parameter  $\delta$ changes in the interval $[0,1]$. For a given
radius $R$, the time parameter for forwardly emitted light increases
as $l$ increases from $l=0$ to $l_{max}(R)$. Then the backward
emissions  takes place for $l < l_{max}(R)$ and ends with $l_T(R)$
when $\delta =1$.

For the radial rays ($l=0$) emitted  from `a' and `b' in Fig. 1 the
time parameter $\delta$ are $0$ and $\delta_f$, respectively. Denote
by $\delta_T$ the time parameter for the point `c' (the last ray
from $R_i$, see Fig.1). Then during the interval
$[\delta_f,\delta_T]$ the distant observer receives light emitted in
the radius domain $[R_i,R_f]$. In other time intervals, earlier,
$0\le \delta \le\delta_f$, or later, $\delta_T\le \delta \le 1$,
only part of this radius domain contributes. For this reason it is
natural to expect the maximum flux at some $\delta$ between
$\delta_f$ and $\delta_T$.

\begin{figure}[htp]
\begin{center}
\includegraphics[height=1.8in,width=1.8in]{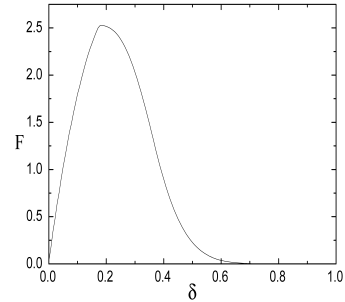}
\caption{The flux from $R_i=6.0M$ through $R_f=4.6M$ in arbitrary
unit.
The horizontal axis is the normalized arrival time $ \delta $. }
\label{flux}
\end{center}
\end{figure}

To illustrate characteristic features of the light curves,  we
plot, in Fig. 2,  the observable flux for the case when the
collapse starts at the radius $R_0=9.0 M$ and the emission  takes
place from $R_i=6.0M$ to $R_f=4.6M$.  We use the analytic
approximation developed in \cite{fl} for calculating  the arrival
time differences, $l \left| dl/ d\tau \right|$ and bending angles.
The arrival time difference between the first ray and the last ray
is calculated to be $\Delta T = 42.0 M$, while  the corresponding
proper time interval is $\Delta \tau = 3.6M$.  The invariant
velocities are $v_I= -0.38$ and $-0.52$ for $R_i$ and $R_f$,
respectively. The first ray from the surface, $R_f$, arrives at
$\delta_f=0.17$ and the last ray from  $R_i$ arrives at $\delta_T
= 0.37$. One can see in Fig. \ref{flux} that the maximum of the
flux is observed around $\delta \sim 0.2$ as discussed above.

\begin{figure}[htp]
\begin{center}
\includegraphics[height=1.8in,width=1.8in]{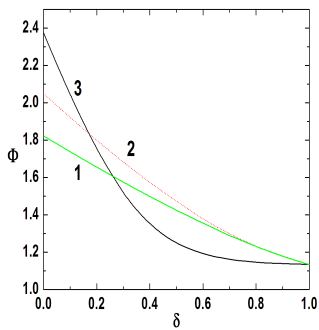}
\caption{The three  lines 1, 2 and 3 are the redshift factors for
the radiation emitted from $R= 6.0M,5.3M$ and $4.6M$ respectively.
Arrival time differences are those from sharp-in-time pulses
\cite{fl}.} \label{redshift}
\end{center}
\end{figure}

Let us discuss now the time dependence of the observed redshift
factor. For a moment, we consider  the radiation of  a sharp-in-time
profile emitted  at the moment of $\tau$ when the radius is $R_e$.
Fig.\ref{redshift} shows the redshift factor as a function of
$\delta$ for $R_e=R_i=6.0 M$, $R_e=5.3M$, and $R_e=R_f=4.6M$. One
can see that  for all three  curves have the same minimum value of
the redshift factor, $\Phi_{T} \equiv \Phi(R,l_T) = 1.13$, at
$\delta=1$.  As discussed in Ref.\cite{fl}, the basic reason is that
the redshift factor of the last ray ($\l=l_T$) for a freely falling
surface does not depend on the radius of radiation but depends only
on the initial radius $R_0$: \ba \label{RST} \Phi_T =
\frac{1}{\sqrt{1-2M/R_0}}\,. \ea

 For the
emission during the {\em finite} interval of time, the lights rays
which arrive at the time $\delta$ are emitted at different radii,
and hence have different redshifts. For each value of $\delta$
(except for $\delta=0$ and $\delta=1$) one has a finite range of
possible redshifts and the redshift curves are broaden.  The
spectral broadening of the monochromatic radiation is  a function of
arrival time.

\begin{figure}[htp]
\begin{center}
\includegraphics[height=1.8in,width=1.8in]{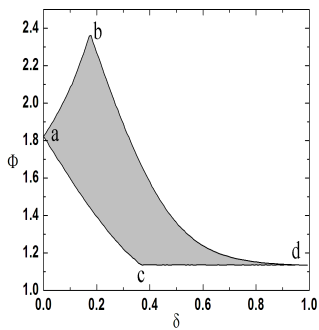} \caption{The
broadening of red shift factors. `a,b,c' and `d' are those in Fig.
\ref{bangle}} \label{rainbow} \end{center} \end{figure}

The shadowed region in Fig. \ref{rainbow} shows the broadening of
redshift factor as a function of the time parameter  $\delta$.  We
again choose $R_0=9M$. The redshift factor of the first ray
(emitted from `a' in Fig. \ref{bangle})   is  $\Phi_0 \equiv
\Phi(R=6.0M,l=0)=1.82 $. The maximum redshift occurs for the ray
emitted from `b' in Fig. \ref{bangle},  which arrives to the
distant observer at $\delta_f=0.18$. The maximum redshift factor
$\Phi_{max} \equiv \Phi(R=4.6,l=0)= 2.36$ in Fig. \ref{redshift}
corresponds to the spectral broadening at $\delta_f$ in Fig.
\ref{rainbow}. From $\delta_T=0.37$ to $\delta=1$, the distant
observer receives the last rays, which are emitted tangentially
for given radii,  with the same redshift factor, $\Phi_{T}$. Hence
the spectral broadening for $\delta_T \leq \delta \leq 1$ has a
constant minimum value of $\Phi_{T}$ between $c$ and $d$, as shown
in Fig. \ref{rainbow}.

In our model, the physical parameters characterizing the collapsing
object  are its mass $M$ and three of  dimensionless parameters
$R_0/M$, $R_i/M$ and $R_f/M$, which determine the initial radius and
the initial and final radiation emitting radii, respectively. In
case we do not know the frequency of the emitted radiation $\nu$ we
cannot determine  the redshift factor directly by observing the
spectrum of the radiation. Nevertheless, if it is possible to
determine the frequency $\nu^{(o)}_{last}$ of the last ray with
sufficient accuracy, then the relative (normalized) redshift factors
$ \tilde{\Phi} \equiv \Phi / \Phi_T= \nu^{(o)}/\nu^{(o)}_{last}$ can
be determined.  Then by  measuring $\delta_f$ and $\delta_T$ as well
as  $\tilde{\Phi}_0$ and $ \tilde{\Phi}_{max}$ one can  determine
${R}_0/M, {R}_i/M$ and ${R}_f/M$. Once we know $R_0/M$, $\Phi_T$ in
Eq.(\ref{RST}) can be evaluated  and  we can determine the original
frequency of emission as given by  $\nu = \Phi_T ~
\nu^{(o)}_{last}$. The mass of the collapsing object can be inferred
from the observed value of $\Delta t$ to complete the determination
of the physical characteristics of the collapsing object.

To summarize, we  discuss in this work the light curves and the
spectral broadening of the radiation  emitted from a surface of a
collapsing object. We demonstrate that the spectral broadening
occurs when the radiation from the collapsing surface takes place
during a finite duration of time.  It is because of the variance of
the frequency shifts of the light rays subjected to  the
gravitational redshift and the Doppler shift of the collapsing
surface. In a simplified model of monochromatic radiation  from a
freely falling spherical surface,  we discuss the possible way how
to infer the physical parameters, the mass, the radii of the
emission and the frequency of the radiation  from the light curves
and spectral broadenings. \vskip .5cm

\noindent The work of V.F. was supported by NSERC and the Killam
Trust. HKL thanks Valeri Frolov for the kind hospitality during
his visit to University of Alberta. HKL was supported by grant No.
(R01-2006-000-10651-0) from the Basic Research Program of the
Korea Science \& Engineering Foundation. A part of this work has
been done during the APCTP-TPI Meeting on {\it Gravity, Cosmology,
and Astrophysics II}, Dec 18-23, 2006, Edmonton, Canada.

\end{document}